\documentclass[%
 aip,
 amsmath,amssymb,
 reprint,%
floatfix
]{revtex4-1}

\usepackage{graphicx}% Include figure files
\usepackage{dcolumn}% Align table columns on decimal point
\usepackage{bm}% bold math
\usepackage[utf8]{inputenc}
\usepackage[T1]{fontenc}
\usepackage{mathptmx}
\usepackage{etoolbox}
\usepackage{textcomp, gensymb}
\usepackage{hyperref}
\usepackage{float}         
\hypersetup{
    colorlinks=true,
    linkcolor=blue,
    filecolor=magenta,      
    urlcolor=blue,
}           

\makeatletter
\def\@email#1#2{%
 \endgroup
 \patchcmd{\titleblock@produce}
  {\frontmatter@RRAPformat}
  {\frontmatter@RRAPformat{\produce@RRAP{#1\href{mailto:#2}{#2}}}\frontmatter@RRAPformat}
  {}{}
}%
\makeatother
\begin{document}

%\preprint{AIP/123-QED}

\title{An Analysis of Muon Flux from Angle Variation of the QuarkNet Cosmic Ray Detector}%[Sample title]
% Force line breaks with \\
\author{Ricco C. Venterea}
 \affiliation{Department of Astronomy, Cornell University, Ithaca, NY, 14853 USA}%Lines break automatically or can be forced with \\ %[Also at ]
 \email{rcv38@cornell.edu}
\author{Urbas Ekka}%
    \affiliation{
    School of Mathematics, University of Minnesota, Twin Cities, MN, 55414 USA%\\This line break forced with \textbackslash\textbackslash
    }%

% \homepage{http://www.Second.institution.edu/~Charlie.Author.}
%\affiliation{%
%Second institution and/or address%\\This line break forced% with \\
%}%

\date{\today}% It is always \today, today,

\begin{abstract}
We present one of the first cosmic ray muon flux-angle variation experiments on the QuarkNet Cosmic Ray Detector (QNCRD). We first describe QNCRD and its calibration. The main focus is then quantifying muon flux decrease as a function of angle from the zenith. The angle of counters of QNCRD were incremented 15\degree \ on average every $3.1$ days over the range of 0\degree \ to 90\degree \ for a period of approximately one month. Results showed that as the angle of the detector increased from the zenith, muon flux decreased, which agrees with previous studies. An estimate for the flux based on the model $I(\theta)=I_0cos(\theta)^n$ had an exponent value of $n=1.39 \pm 0.01$ for $\theta \leq 75 \degree$, an underestimate of values in other literature. These findings provided a reasonable, although not entirely accurate, estimate for the value of $n$ considering the duration of the study and sensitivity of the instrument. Our results constrain the accuracy of QNCRD and provide a source for future long-term experiments. This study also demonstrates the feasibility of conducting science experiments in high school classrooms, increasing science accessibility. 
\end{abstract}

\maketitle

%\begin{quotation}
%The ``lead paragraph'' is encapsulated with the \LaTeX\ 
%\verb+quotation+ environment and is formatted as a single paragraph before the %%first section heading. 
%(The \verb+quotation+ environment reverts to its usual meaning after the first %sectioning command.) 
%Note that numbered references are allowed in the lead paragraph.
%%
%The lead paragraph will only be found in an article being prepared for the %journal \textit{Chaos}.
%\end{quotation}

\section{\label{sec:Introduction}Introduction}

QuarkNet is part of a National Science Foundation funded effort to increase science accessibility across high school classrooms in the United States. This effort also includes training for high school science teachers and students.\cite{Bardeen2018} As part of this network, QuarkNet Detectors are located throughout the world in high school classrooms.\cite{QuarknetHS} These detectors have been used to create and study particle physics experiments in classrooms, which range from the impact of solar eclipses on cosmic ray muon flux\cite{solar_eclipse} to determining average zenith muon flux rate.\cite{Shaffer_Quark}

Muons are a byproduct of cosmic rays, a stream of particles constantly entering Earth's atmosphere. Cosmic rays are composed of highly energetic particles, mostly consisting of hydrogen and helium nuclei. High-energy cosmic rays originate from neutron stars, while low-energy rays originate from the Sun.\cite{Gaisser2016} The QuarkNet Detector measures\cite{solar_eclipse} muon flux momenta greater than $2$ GeV. %The energy of these rays can range from $10^9$ eV to beyond $10^{20}$ eV \cite{energy_rays}. 

When cosmic rays enter Earth's atmosphere, they collide with air molecules, creating a cascading effect known as an air shower.\cite{Gaisser2016} After this collision, pions are produced, some of which decay into muons. Other pions continue into earth's atmosphere and interact with air molecules, creating more air showers.\cite{rao_extensive_1998} Muons are similar to electrons, with a negative charge and about 200 times as massive than the electron.\cite{constants}

These muons can be measured using detectors on earth. As the muons pass through a scintillation counter, they interact with electrons, which release photons. These photons are reflected inside the counter until they reach a detector, where they are transformed into electric signals.

It is currently known that muon flux decreases as the angle of a muon detector increases from the zenith. Previous muon flux studies have used more precise detection methods at various latitudes, longitudes, and altitudes. Shukla and Sankrith\cite{shukla_energy_2018} describe the theoretical and experimental flux values for a muon detector located at sea level based on the $\cos^2{\theta}$ model. They also implement their own best-fit model, $I(\theta) = I_0 D(\theta)^{-(n-1)}$, with $n = 3.09 \pm 0.03$. We also implement this model. Schwerdt\cite{schwerdt_zenith_2018} presents a model based on $I(\theta) = a\cos^2{(b\theta + c)} + d$. Pethuraj et al.\cite{Pethuraj_2017} model muonic flux as a function of angle and arrive at an exponential value of $n = 2.00 \pm 0.04$ for their model $I(\theta) = I_0\cos^{n}{\theta}$, located $\approx 160$ m above sea level. The cos-squared model is the main comparison of this study, but we also present values for the Schwerdt,\cite{schwerdt_zenith_2018} Shukla,\cite{shukla_energy_2018} and Pethuraj et al.\cite{Pethuraj_2017} models. We find the Schwerdt\cite{schwerdt_zenith_2018} model provides the best fit for our data compared to the aforementioned models. 

\begin{figure}[h]
    \centering
    \includegraphics[scale=0.6]{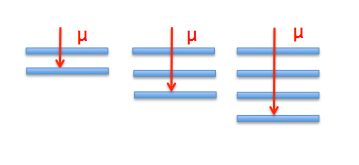}
    \caption{Illustration of two-, three-, and four-fold coincidence, where the muon is only detected if it passes through all counters in each coincidence configuration.\cite{eLab}}
    \label{fig:Coincidence}
\end{figure}

While not a flux experiment like the one presented in this paper, Shaffer presents average muon flux rate results using the QuarkNet Detector. Shaffer used QNCRD at an angle of $0 \degree$ from the zenith to measure muon flux near Topeka, Kansas and found a flux rate of $1200$ to $1500$ events per meter squared per minute per steradian, values significantly lower than those found in this study. Shaffer's plateau values were different than those used in this study and collected data for several weeks, while this study was one month of data. Shaffer presents a novel solution to measuring steradians using the QuarkNet Detector, a conversion we use in this study. Shaffer's total detector distance between top and bottom counter was $40$ cm total, while the maximum spacing for our counters was $13$ cm. The coincidence rate used in the plateauing process in this study is approximately equal to the rate used by Shaffer.\cite{Shaffer_Quark} Coincidence rate is the number of counters needed to qualify muon signals as a detection (see Figure \ref{fig:Coincidence}).

In Section \ref{sec:Quarknet Detector}, we discuss features of the QuarkNet Detector and provide a description of the calibration process of QNCRD. The experiment is described in Section \ref{sec:Experiment}. Results from this flux experiment can be found in Section \ref{sec:Results}, followed by a data analysis in Section \ref{sec:Analysis}. We follow the analysis with a discussion in Section \ref{sec:Discussion}. We conclude with relevant findings and further improvements in Section \ref{sec:Conclusions}. 

\section{Quarknet Detector}\label{sec:Quarknet Detector}

The detector used in this study is located at Irondale Senior High School, with coordinates $45.0900\degree N, 93.2072\degree W$ at an altitude of $276$ meters. 

QNCRD consists of a data acquisition (DAQ) board, four scintillation counters, Equip software, photomultiplier tubes (PMTs), Global Positioning System (GPS) receiver, a power supply and a power distribution unit (PDU). Each plastic counter has dimensions $25.4$ cm x $30.5$ cm x $1.27$ cm and a "cookie" attached to one corner.\cite{QUACRD_instructions} This cookie is the interface between the counter and the photomultiplier tube. Each counter is wrapped in reflective shielding to retain any signal from muon interactions. Counters will also be referred to as channels throughout this paper, being labeled as channel $0$, channel $1$, channel $2$, and channel $3$.

Each counter has a photomultiplier tube (PMT) that collects electric signals as muons pass through the shielding.\cite{lofgren_quarknet_2001} The PMTs are SensTech Model P30CW5 photodetector packages.\cite{QUACRD_instructions} These photodetectors are connected to a power supply\cite{Shaffer_Quark} and controlled by a PDU, with voltages in the range $0.30 $ V to $5.0$ V.

The data acquisition board is the circuitry necessary for collecting electric signals via the PMTs. This data board also generates the output data to be uploaded to the Equip software.\cite{QUACRD_instructions} There is a 1.25 ns resolution on the data board, used for separating muon events and unrelated and unwanted ion events. \cite{Shaffer_Quark}

The GPS antenna is placed outside the school building at all times, while the GPS box is located inside, next to the data acquisition board. The GPS provides the location of the instrument as well as times of the events, which is accurate\cite{Shaffer_Quark} to $24$ ns.

The Equip software records the channels in use, as well as temperature, location, and flux. This also includes the coincidence rate. The software provides an interface to the Cosmic Ray e-Lab, a website where data from QuarkNet Cosmic Ray Detectors across the world are uploaded.\cite{eLab} Equip was installed and ran on a Windows XP operating system.

\subsection{\label{sec:Plateau Process}Plateau process}

To ensure reliability of data, performance studies are conducted to measure the time over threshold (ToT) of the PMT to a muon event.\cite{performance} The ToT is defined as the amount of time an event is above a predetermined threshold level.\cite{signal} Without this process, the flux may be an under- or overestimate of the true value of muons passing through the counters. Such inaccuracies would be due to a high or low voltage value of the PMTs. Voltage values are adjusted through the power supply. 

The plateau process involves setting the power supply to $0.3$ V, the lowest voltage setting. In order to plateau one counter, another counter has to be used as a reference. Counters 0 and 1 were stacked, channel 0 serving as a reference. The threshold level of the three detectors was set to 300 mV by typing \textit{TL 4 300} into the Equip software. Channel 1 was activated and read a one-fold coincidence. One-fold coincidence means that a muon needs to travel through one detector to be counted as a detection (Figure \ref{fig:Coincidence}). Waiting for $10$ seconds, the voltage was increased until the digital counter on the DAQ board was between $400$ to $600$ counts. Once the counter was within this range, the voltage was gradually increased until the coincidence counts levelled off.  
%The purpose of the plateau process is to calibrate the detectors so they can accurately measure the flux of muons passing through the counters. 

\begin{figure}[h]
    \centering
    \includegraphics[scale=0.65]{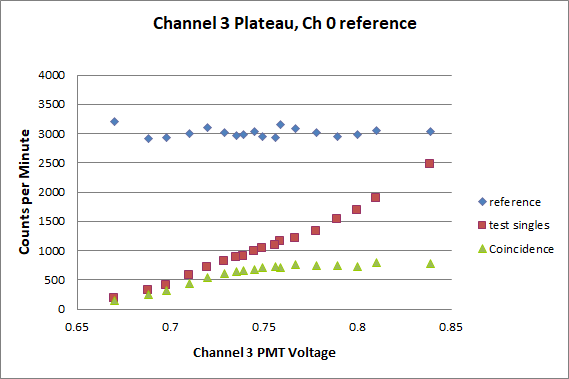}
    \caption{Plateauing coincidence rate in channel 3 is indicated by the green triangles. The coincidence of channel 3 plateaued within a specific voltage range. This plateauing is from the second round of calibration.}
    \label{fig:PlatSucc}
\end{figure}

This process was repeated for counters 2 and 3, channel 0 serving as the reference channel.\cite{paschke_calibration_2009} Results for channel 3 are shown in Figure \ref{fig:PlatSucc}. After this plateau process, data were collected over the next two weeks for eight hours each day. This data was collected to ensure the counters were calibrated correctly. The counters were stacked 1, 0, 2, 3, from bottom to top, with a 0 degree angle from the horizon. Three-fold coincidence was used for counters 0, 2, and 3, as counter 1 was determined inoperable due to improper wrapping of the reflective shielding in that counter. 

During the initial plateau process, results from counter $2$ started to cause concern. The channel rate was displaying a peak pattern at the same time of day. This prompted a re-calibration of QNCRD as the preliminary voltage was set too low. The plateau process was therefore repeated, with new data collected over two weeks. The voltages after the second plateau process are shown in Table \ref{table:Voltage}, which were used for the remainder of this study. The voltage values were estimated from the plateau graphs where the coincidence values were just beginning to plateau (Figure \ref{fig:PlatSucc}).

\begin{table}[h]
    \centering
    \caption{Voltage values determined from plateau process. See Figure \ref{fig:PlatSucc} for the voltage estimate in counter 3.}
    \begin{tabular}{|c|c|c|} 
        \hline
        Counter & Voltage (V)\\
        \hline
        3 & 0.770 \\ 
        \hline
        2 & 0.800 \\
        \hline
        0 & 0.709 \\
        \hline
    \end{tabular}
    \label{table:Voltage}
\end{table}

%The ideal plateau results are shown in Figure \ref{fig:PlatSucc}.  

%The channels were plateaued a second time following Adams' advice. 

%Using the new voltage values in Table \ref{table:Voltage}, data were again collected for two weeks at a 0\degree \ angle. The peaks had vanished from subsequent data sets, and no other patterns were observed.

%As can be seen, the coincidence of channel 1 stayed relatively constant.  It is believed that this was a result of incorrect wrapping of the counter, which is the author's fault.  This counter was never used in the experiment.      

%\begin{figure}[h]
%    \centering
%    \includegraphics[width=\linewidth,scale=0.25]{Images/PlateauFail.png}
%    \caption{No plateauing in coincidence rate of channel 1.  Coincidence is the signals received within a short time determined by the plateau process \cite{noauthor_cosmic_nodate}.  Test singles is the rate at which one counter is being hit by muons \cite{rylander_quarknet_2010}.}
%    \label{fig:PlatFail}
%\end{figure}

\section{\label{sec:Experiment}Experiment}

Data was collected almost every day between the end of October 2019 to the beginning of December 2019. The initial angle of the detectors was $0$ degrees from the horizon (i.e., parallel to the horizon). The detector angle was incremented by $15$ degrees approximately every $3.1$ days, with the study ending with the panel surfaces perpendicular to the horizon, defined to be $90$ degrees. Data was uploaded from the QuarkNet Detector to the Cosmic Ray e-Lab website, where raw flux data was extracted. 

\begin{figure}[h]
    \centering
    \includegraphics[scale=0.2]{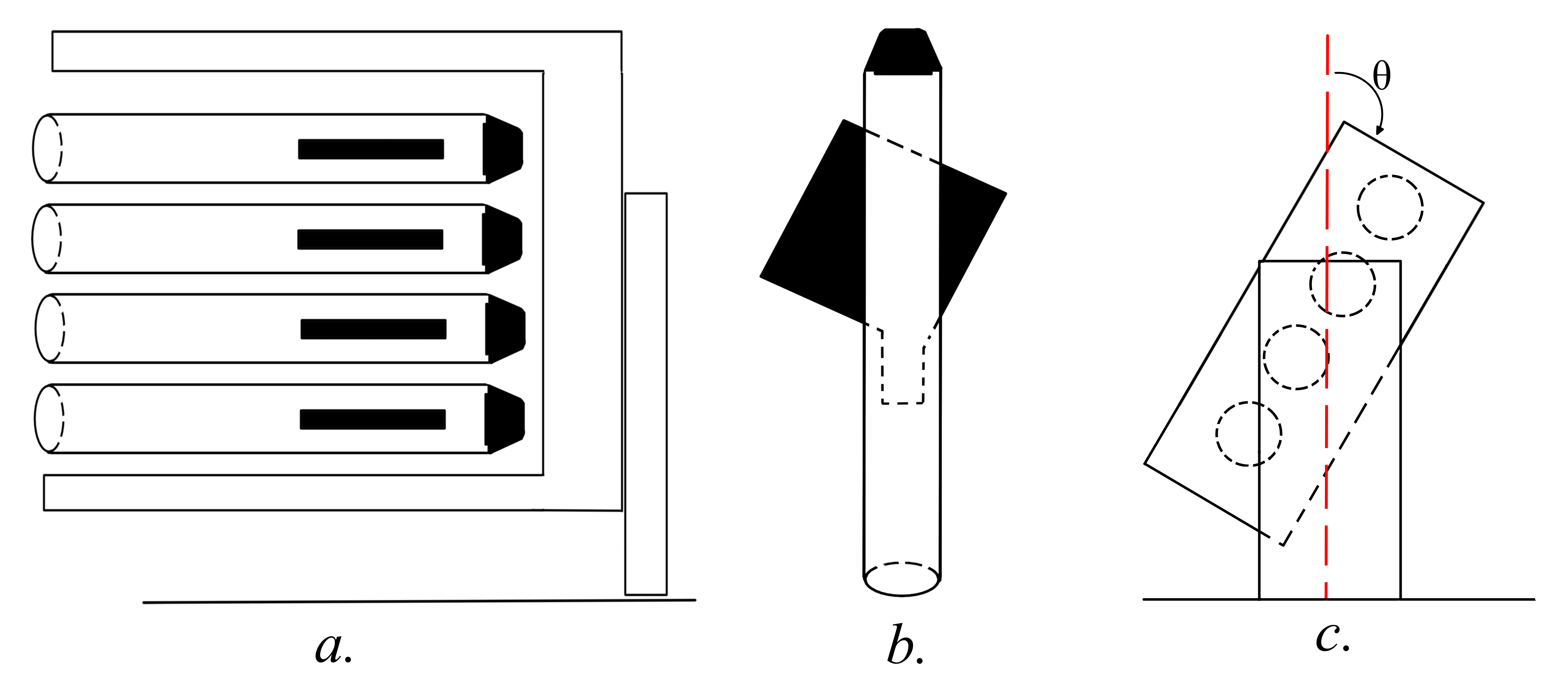}
    \caption{Schematic of the detector configuration is highlighted in Figure \ref{fig:diagram}$a$. Starting from the top, the counters are labeled Counter $3$, Counter $2$, Counter $0$, and Counter $1$. The individual counters are shown in Figure \ref{fig:diagram}$b$ and are represented as black rectangles. Figure \ref{fig:diagram}$c$ represents how the QuarkNet Detector angle was varied throughout the duration of this experiment. Note that only the top three counters were used in this study, as the fourth detector was deemed inoperable. Diagrams are not to scale.}
    \label{fig:diagram}
\end{figure}

\begin{table}[h]
    \centering
    \caption{Relative vertical spacing of counters used in this study, with negative values being measured below the origin set at the GPS box.}
    \begin{tabular}{|c|c|} 
        \hline
        Counter & Vertical Spacing (m) \\
        \hline
        3 & -1.5  \\ 
        \hline
        2 & -1.565 \\
        \hline
        0 & -1.63 \\
        \hline
    \end{tabular}
    \label{table:Spacing}
\end{table}

The QuarkNet Cosmic Ray Muon Detector was set up using three counters (see Figure \ref{fig:diagram}) and three-fold coincidence. The configuration of the counters remained stacked 1, 0, 2, 3, from bottom to top. The relative vertical spacing measured from the QNCRD GPS box is provided in Table \ref{table:Spacing} and illustrated in Figure \ref{fig:diagram}.

%After the plateau process (described below), counter 1 was found to have a possible light leak, so it was not used for the rest of the experiment.  A light leak prevents accurate data collection as electrons can escape from the counter. While the Quarknet detector is composed of four detector panels, only three were operational due to a malfunction in one.

\section{\label{sec:Results}Flux Results}

%\subsection*{Before Voltage Change}
%Three counters were used to collect data at a 0\degree \ angle.  This meant that the counters were parallel to the surface of the Earth.  Results showed that channel 2 seemed to have a noticeable peak every day, occurring between 12 - 15 UTC.  This is shown in Figure \ref{fig:Anomaly}.  Voltage changes to the counters fixed this anomaly.

%\begin{figure}[h]
%    \centering
%    \includegraphics[width=\linewidth,scale=0.25]{Images/AnomalyChannels.jpg}
%    \caption{The Cosmic Ray e-Lab, which was used to generate this graph, refers to channels as 1, 2, 3, 4, whereas the equipment refers to them as counter 0, 1, 2, 3.  Counter 1 was not used in the study.  Mysterious peaks in counter 2 consistently occurred at the same time of day.}
%    \label{fig:Anomaly}
%\end{figure}

%\subsection*{After Voltage Change}
The flux data were collected at 15\degree \ increments for a minimum of 1.5 days using the voltage values in Table \ref{table:Voltage}. The detector's final angle ended at 90\degree \ (a vertical orientation relative to the horizon). Data were not collected on Wednesdays because computers in the school were automatically turned off, preventing any data collection during this time period. 

Since the QuarkNet Detector does not have the capability of measuring flux per steradian, we use Equation \ref{eqn:tan} as described by Shaffer to convert our data to make its analysis easier:\cite{Shaffer_Quark}

\begin{equation} \label{eqn:tan}
    \tan{\theta} = \frac{w}{d}
\end{equation}

where $w$ is the width of the detector and $d$ is the total distance between the top and bottom channel. Utilizing the fact that the angular measurement of $32.77$ degrees from the normal equals one steradian, the width of the QuarkNet Detector counter being $0.26$ m, and the distance between the top and bottom counter of the detector being $0.13$ m (see Table \ref{table:Spacing}), we found the needed adjustment of data to be:\cite{steradian}

\begin{equation*}
    \arctan{(\frac{w}{d})} \cdot \frac{1 sr}{32.77 \degree} = \arctan{(\frac{0.26 m}{0.13 m})} \cdot \frac{1 sr}{32.77 \degree} = 1.936 sr
\end{equation*}

\begin{figure}[h]
    \centering
    \includegraphics[scale=0.5]{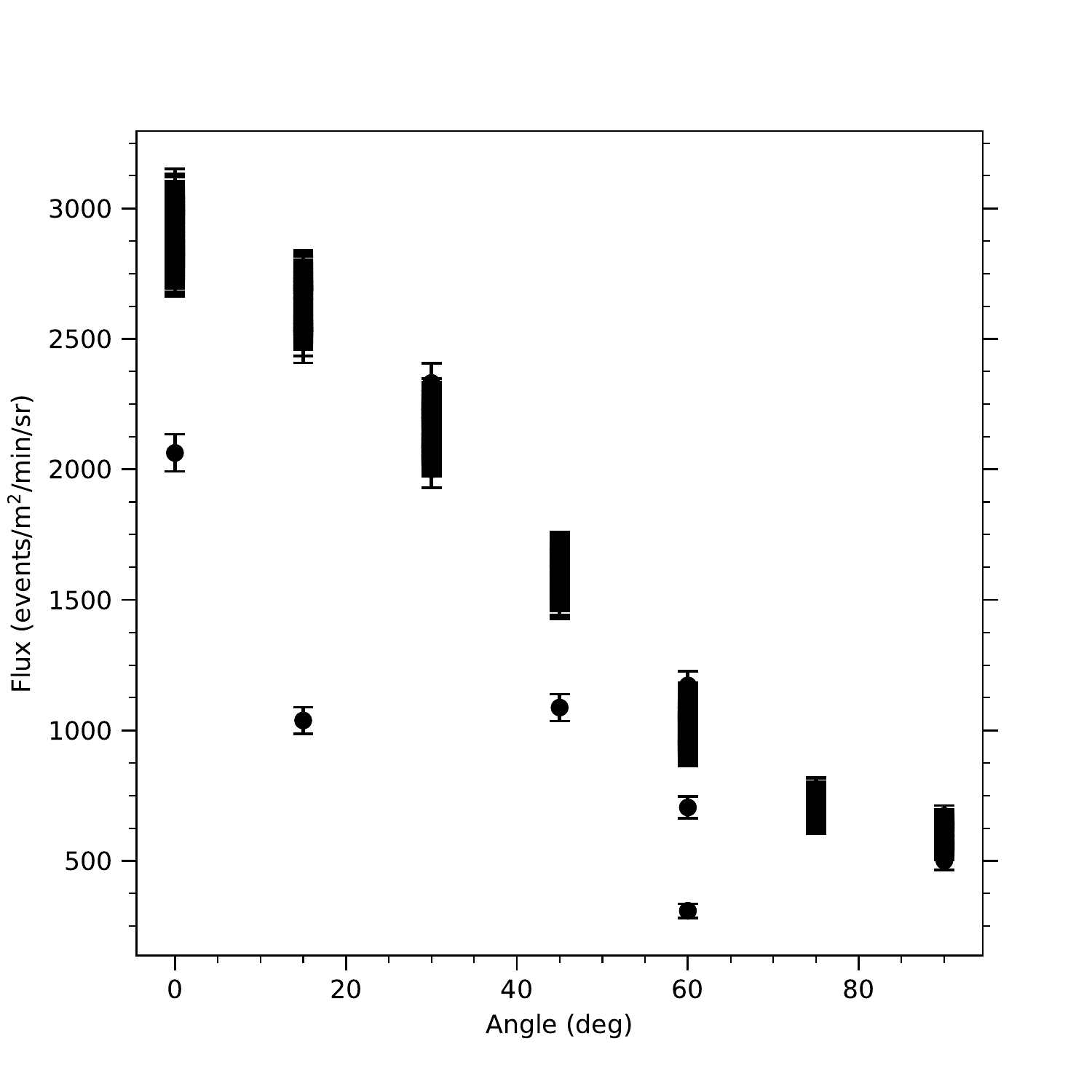}
    \caption{Decrease of flux starting at 0\degree \ ending at 90\degree. These results are from channel $0$ for detector $6709$.}
    \label{fig:All}
\end{figure}

We now converted flux results from the Cosmic Ray e-Lab into units of $events/m^2/min/sr$. A combination of all the muon flux measurements at increasing detector angles is shown in Figure \ref{fig:All}. This experiment continues to verify the general trend that muon flux decreases as the angle of the detector increases from the zenith. 

While collecting data, we found discrepancies in data at $30$\degree \ and $75$\degree. We attributed this to an improper alignment of the counters and a computer malfunction. We resolved these issues by conducting the experiment again over $1.5$ days for each affected angle.

\section{\label{sec:Analysis}Analysis}

A statistical analysis was performed on the flux data, which included removing outliers and then comparing data to the $\cos^2{\theta}$ function, as this is widely believed to be the most accurate description of muonic flux as a function of angle.\cite{shukla_energy_2018}$^,$ \cite{Pethuraj_2017}$^,$ \cite{Shaffer_Quark}$^,$ \cite{bektasoglu_investigation_2013}$^,$ \cite{shteinbuk_measuring_nodate} We also fitted flux data to models given by Shukla and Sankrith,\cite{schwerdt_zenith_2018} Schwerdt,\cite{shukla_energy_2018} and Pethuraj et al.\cite{Pethuraj_2017} We find Schwerdt's\cite{schwerdt_zenith_2018} model best represents the data presented in this paper, which is supported by a reduced chi-squared test with a value of $\chi{^2}_{\nu} = 0.467$.

We first cut outliers in the flux data, using the $25$th and $75$th percentile values. These outliers were based on the interquartile range of the data. Any values lying below the $25$th percentile minus $1.5$ times the interquartile range were cut and values above the $75$th percentile plus $1.5$ times the interquartile range were cut. This removes the $5$ outliers as seen in Figure \ref{fig:All}.

Due to construction occurring during measurements, it is possible that electrical interference may have affected the results causing the outliers in our data. Another cause of outliers could have been the relative age of the detector, and according to Bae and Chatzidakis, detections of high zenith angles ($\theta >60$\degree) saw high levels of uncertainty.\cite{Bae} However, we find outliers for $\theta < 60$\degree. 

\begin{figure}[h]
    \centering
    \includegraphics[scale=0.60]{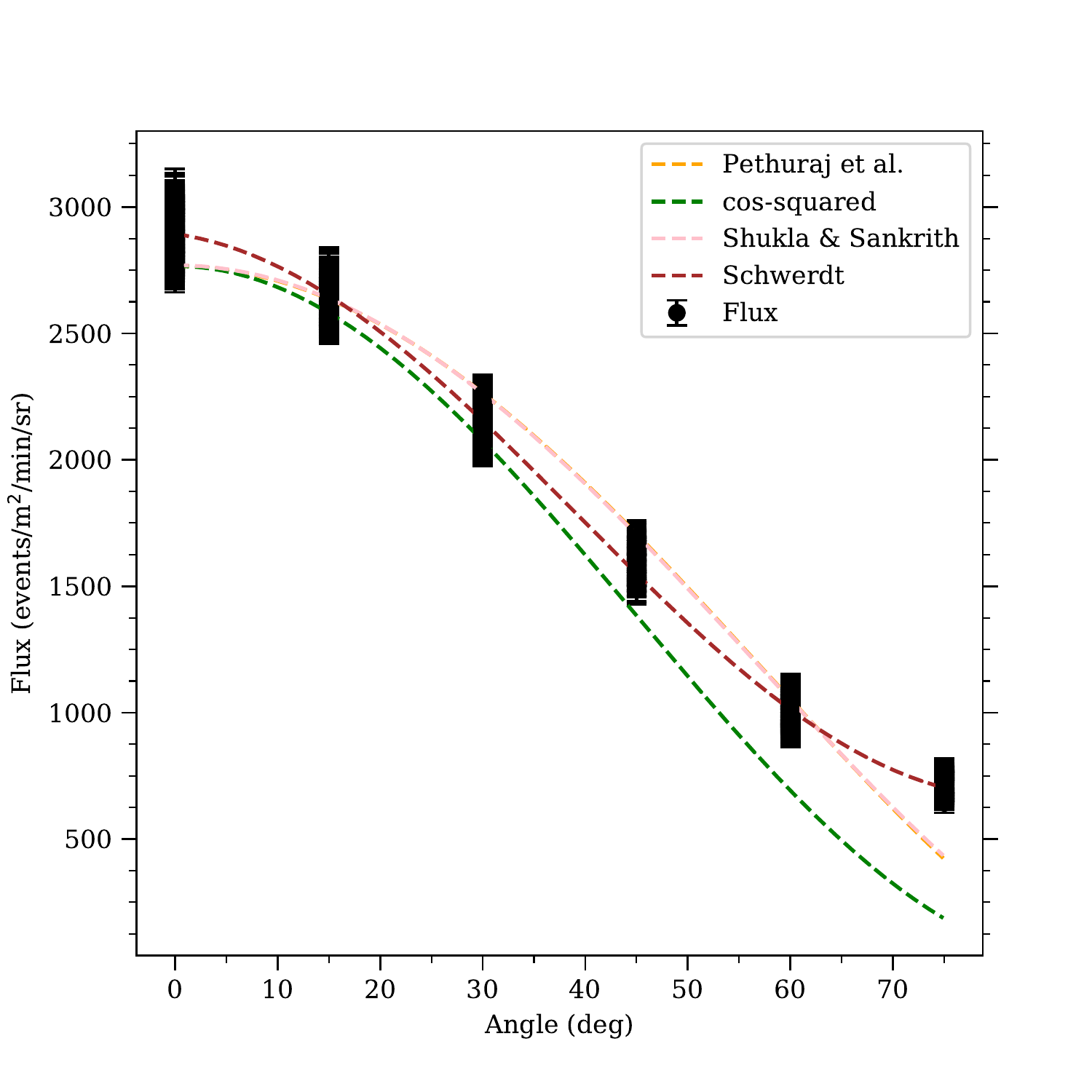}
    \caption{Comparison of muon flux models as a function of angle with data collected in this study. Note that we only fit for $\theta \leq 75 \degree$. Observe how well the Schwerdt\cite{schwerdt_zenith_2018} model agrees with the data. The Shukla and Sankrith\cite{shukla_energy_2018} model and Pethuraj et al.\cite{Pethuraj_2017} model agree so well they overlap.}
    \label{fig:flux_comp}
\end{figure}

Once we cleaned the data of outliers, we were able to generate a comparison based on the $cos^2(\theta)$ model. Converting the flux in units of steradians also helps validate our findings, as the data can now be easily compared to other flux studies. This comparison is seen in Figure \ref{fig:flux_comp}. 

Grieder and Pethuraj et al. state that the intensity of muons follows the empirical model:\cite{grieder_cosmic_2001}$^,$ \cite{Pethuraj_2017}

\begin{equation}    \label{General}
 I(\theta)=I_0\cos^n{\theta}
\end{equation}

where $I_0$ is the vertical intensity and $\theta$ is the angle from the zenith. However, this equation is used to approximate intensities\cite{grieder_cosmic_2001} only for $\theta \leq 75\degree$. Using the curve\_fit function from the scipy.optimize package and the experimentally determined Pethuraj et al.\cite{Pethuraj_2017} model, we fit our data for $\theta \leq 75$\degree. We calculated an exponential value of $n = 1.39 \pm 0.00657$, with $r^2 = 0.973$.

We also try modeling to the Shukla and Sankrith function, given by:\cite{shukla_energy_2018}

\begin{equation} \label{eqn: Shukla}
    I(\theta) = I_0 D(\theta)^{-(n-1)}
\end{equation}

with

\begin{equation} \label{eqn:D_func}
    D(\theta) = \sqrt{(\frac{R^2}{d^2}\cos^2{\theta} + 2\frac{R}{d} + 1)} - \frac{R}{d}\cos{\theta}
\end{equation}

where Shukla and Sankrith\cite{shukla_energy_2018} fit the ratio $R/d = 174.0$. Using this model, we find $n=2.41 \pm 0.00645$ with $r^2 = 0.974$. 

Finally, we compare to the Schwerdt model:\cite{schwerdt_zenith_2018}

\begin{equation} \label{eqn:schwerdt}
    I(\theta) = a\cos^2{(b\theta + c)} + d
\end{equation}

with $a$ representing a vertical stretch, $b$ a horizontal stretch, $c$ a horizontal stretch, and $d$ a vertical shift. This model is more mathematical in nature, representing the most general form of the cos-squared function.\cite{schwerdt_zenith_2018} We find this model best represents the data, especially towards increasing values of $\theta$ and has $r^2 = 0.997$. This function is approximated by 

\begin{equation} \label{eqn: schwerdt_fit}
    I(\theta) = 2241.21 \cos^2{(1.047 \theta + 0.0678)} + 665.13
\end{equation}

There are several differing values for $n$, all looking at flux data with $\theta \leq75\degree$. Grieder\cite{grieder_cosmic_2001} states that the average value of $n=1.85 \pm0.10$. Useche and Avila\cite{useche_estimation_2019} state the experimental value for $n=1.96 \pm 0.22$. Other results\cite{bektasoglu_investigation_2013}$^,$\cite{useche_estimation_2019} have estimated the value of $n$ to be $n=1.95 \pm 0.08$ and $n=2.11 \pm 0.03$. See Figure \ref{fig:flux_comp} for a comparison of models mentioned in this paper with flux data collected in this study. See also Table \ref{tab:studies} for a comparison of exponential values from other studies.

{}

\begin{table}[h]
    \centering
    \begin{tabular}{|c|c|c|c|c|}
        \hline
         Authors & Mag. Lat. (\degree N) & Alt. (m) &  \emph{n} value \\
         \hline
         Crookes and Rastin & $53$ & $40$ & $2.16 \pm 0.01$  \\
         \hline
         Greisen & $54$ & $259$ & $2.1$ \\
         \hline
         Judge and Nash & $53$ & $0$ & $1.96 \pm 0.22$ \\
         \hline
         Karmakar et al & $16$ & $122$ & $2.2$ \\
         \hline
         S.Pal & $10.61$ & $0$ & $2.15 \pm 0.01$ \\
         \hline
         S. Pethuraj et al. & $1.44$ & $160$ & $2.00 \pm 0.04$ \\
         \hline
         \emph{This study} & $45$ & $276$ & $1.39 \pm 0.01$ \\     
         \hline
    \end{tabular}
    \caption{Comparison of muon flux data to previous studies.\cite{Pethuraj_2017} Note that $0$ m altitude corresponds to sea level.}
    \label{tab:studies}
\end{table}

We also perform a chi-squared test on the four models discussed in this study. We find that the reduced chi-squared $\chi^2_{\nu}$ value for the Schwerdt\cite{schwerdt_zenith_2018} model best represents the data, which is also supported by a coefficient of determination of $r^2 = 0.997$. Figure \ref{fig:flux_comp} displays how well this model follows the data. We summarize our statistical results in Table \ref{tab:stats}.

\begin{table}[h]
    \centering
    \begin{tabular}{|c|c|c|c|}
        \hline
         Model & Equation &  $\chi^2_{\nu}$ & $r^2$  \\
         \hline
         Pethuraj et al. & $I_0\cos^n{\theta}$ & $5.70$ & $0.973$ \\
         \hline
         Shukla &  $I_0 D(\theta)^{-(n-1)}$ & $5.35$ & $0.974$ \\
         \hline
         Schwerdt & $a\cos^2{(b\theta + c)} + d$ & $0.467$ & $0.99$7 \\
         \hline
         cos-squared & $I_0\cos^2{\theta}$ & $22.7$ & $0.916$ \\
         \hline
    \end{tabular}
    \caption{Summary of statistical tests performed on the four models presented in this study for $\theta \leq 75 \degree$.}
    \label{tab:stats}
\end{table}

\section{\label{sec:Discussion}Discussion}

In general, we saw the flux decrease as the angle increased, which agrees with measurements made by Schwerdt\cite{schwerdt_zenith_2018} and Useche and Avila.\cite{useche_estimation_2019} The reason for this flux decrease is that as the angle increases, the cosmic ray muons are not able to penetrate the counters at extreme angles. The majority of muons are ``raining" down on the channels at 0\degree; the intensity of muons entering at 90\degree \ would be significantly less.  

The improved model to estimate the cosmic ray muon flux for all detection conditions is especially significant for high zenith angles ($\theta > 60$\degree) because the cosine-squared model is limited in use for low zenith angles due to large uncertainties and assumes a flat earth model.\cite{Bae}$^,$ \cite{Olmos_Y_ez_2021} 

While the reduced chi-squared test favored the Schwerdt model, a value of $\chi^2_{\nu} \approx 1$ suggests the model and measurements follow error variance. For the Schwerdt model, we found the reduced chi-squared to be slightly less than $1$, which may suggest an improper error fit for this model.\cite{schwerdt_zenith_2018} 

Our results highlight a significant underestimate of previous flux studies, with most agreeing with $n = 2$ for the exponential value. However, given the duration of this study, as well as the equipment used, this presents a precise, although not entirely accurate, flux study. Indeed, if the duration of the study were longer, more flux data could be obtained, presenting more values to use in best fit models. Additionally, only three counters were in full operation. Using a fourth counter would change the flux data being collected, since cosmic ray muons would now have to traverse four counters in order to be classified as a detection (Figure \ref{fig:Coincidence}). However, this would add an additional distance of approximately $10$ cm to the top and bottom detector distance, which may change the flux data. In this case, it would be expected for the cosmic ray muon flux to decrease since a fourth counter would raise the requirement of being considered a measurement. The location of the detector may also explain why we found flux values to be below the expected value of $n = 2$. Construction was ongoing throughout the school day (8:35 AM CST to 3:15 PM CST). Any electrical interference may have affected these results. Our results in general suggest an under performance of the QuarkNet Detector.

\section{\label{sec:Conclusions}Conclusion}

We conducted a short-term cosmic ray muon flux experiment to test the cos-squared model using the QuarkNet Cosmic Ray Detector and found the muon flux to decrease as the detector angle from the zenith increased. This agrees with previous experiments that varied the angle of cosmic ray detectors. For relatively low angles ($\theta < 45 \degree$), our results roughly correspond to the $\cos^2{\theta}$ model, with several areas for improvement. We did find discrepancies in flux data at $30 \degree$ and $75 \degree$, which we attributed to experimental issues. Resolving such issues would most likely improve our current flux model. We also found the Schwerdt\cite{schwerdt_zenith_2018} model to best represent the data, especially for data at high angles. The vertical shift in this model accurately accounts for muon flux at larger detector angles, whereas the simple cos-squared model would yield a null flux result.

Our findings are important for several reasons. Most notably, this is one of the first type of experiments performed on the QuarkNet Cosmic Ray Muon Detector to analyze the relation between muon flux and angle. This paper serves as a baseline for future studies that can improve upon our current value of $n = 1.39$ for QNCRD. These results also present an accurate representation of other flux experiments as detailed here, legitimizing the QuarkNet Detector as a tool for scientific research and study. 

While our study was short compared to previous studies, and involved equipment with much less sensitivity than other detectors, these results are important for improving the accuracy of the detector. High schoolers, as well as particle physicists, may use this paper to guide their own studies similar to this. 

This experiment not only provides a constraint on the accuracy of detecting muonic flux at varying angles for QNCRD, but also provides a framework for future long-term studies to be implemented resembling this experiment description. We have several suggestions that could improve the accuracy of this detector. More data should be collected over a longer period of time for several reasons.  Ensuring that there is no seasonal, diurnal, or other temporal variations could not have been completely verified with this short of study. Collecting over a longer period of time would increase the data set, improving flux collection at various angles. We suggest collecting data over at least six months.  

Using a fourth counter should improve the results. Having a four-fold coincidence would greatly increase the accuracy of the data and may bring the flux results down for high angles. 

The effect of latitude on cosmic ray muon flux can be better studied with this detector. Given that QNCRD is located across the globe, this paper shows that a worldwide study may be performed to better understand how flux changes with latitude.\cite{QuarknetHS} A similar study may be performed to study variations in altitude.

%It is still unknown why peaks in counter 2 occurred as a result of the first plateau attempt.  Increasing the voltage values may be an explanation for the disappearance of the peaks.  

\begin{acknowledgements}
We would like to thank funding and support from Fermilab, QuarkNet, the National Science Foundation, the Irondale High School administration, and Mark Adams of Fermilab.

RCV would like to thank Shane Wood for taking the time to meet with him after school and directing this study, as well as providing comments for this paper. From his class, RCV learned the fundamentals of cosmic rays, but with his guidance, he has been able to explore the subatomic world in much more depth. RCV would also like to thank Dr. Mary Sande and Logan Doroff for allowing this study to take place in their high school classroom. Finally, RCV would like to thank Rodney Venterea for supporting his scientific endeavors and providing feedback for this paper.    
\end{acknowledgements}

\section*{Data Availability}
The data used in this study are publicly available here: \url{https://github.com/ricco-hub/cosmic_rays/tree/master/Angles}. 

\section*{References}
\nocite{*}
\bibliography{aipsamp}% Produces the bibliography via BibTeX.

\begin{thebibliography}{23}%
\makeatletter
\providecommand \@ifxundefined [1]{%
 \@ifx{#1\undefined}
}%
\providecommand \@ifnum [1]{%
 \ifnum #1\expandafter \@firstoftwo
 \else \expandafter \@secondoftwo
 \fi
}%
\providecommand \@ifx [1]{%
 \ifx #1\expandafter \@firstoftwo
 \else \expandafter \@secondoftwo
 \fi
}%
\providecommand \natexlab [1]{#1}%
\providecommand \enquote  [1]{``#1''}%
\providecommand \bibnamefont  [1]{#1}%
\providecommand \bibfnamefont [1]{#1}%
\providecommand \citenamefont [1]{#1}%
\providecommand \href@noop [0]{\@secondoftwo}%
\providecommand \href [0]{\begingroup \@sanitize@url \@href}%
\providecommand \@href[1]{\@@startlink{#1}\@@href}%
\providecommand \@@href[1]{\endgroup#1\@@endlink}%
\providecommand \@sanitize@url [0]{\catcode `\\12\catcode `\$12\catcode
  `\&12\catcode `\#12\catcode `\^12\catcode `\_12\catcode `\%12\relax}%
\providecommand \@@startlink[1]{}%
\providecommand \@@endlink[0]{}%
\providecommand \url  [0]{\begingroup\@sanitize@url \@url }%
\providecommand \@url [1]{\endgroup\@href {#1}{\urlprefix }}%
\providecommand \urlprefix  [0]{URL }%
\providecommand \Eprint [0]{\href }%
\providecommand \doibase [0]{http://dx.doi.org/}%
\providecommand \selectlanguage [0]{\@gobble}%
\providecommand \bibinfo  [0]{\@secondoftwo}%
\providecommand \bibfield  [0]{\@secondoftwo}%
\providecommand \translation [1]{[#1]}%
\providecommand \BibitemOpen [0]{}%
\providecommand \bibitemStop [0]{}%
\providecommand \bibitemNoStop [0]{.\EOS\space}%
\providecommand \EOS [0]{\spacefactor3000\relax}%
\providecommand \BibitemShut  [1]{\csname bibitem#1\endcsname}%
\let\auto@bib@innerbib\@empty
%</preamble>
\bibitem [{\citenamefont {Bardeen}\ \emph {et~al.}(2018)\citenamefont
  {Bardeen}, \citenamefont {Wayne},\ and\ \citenamefont {Young}}]{Bardeen2018}%
  \BibitemOpen
  \bibfield  {author} {\bibinfo {author} {\bibfnamefont {M.}~\bibnamefont
  {Bardeen}}, \bibinfo {author} {\bibfnamefont {M.}~\bibnamefont {Wayne}}, \
  and\ \bibinfo {author} {\bibfnamefont {M.~J.}\ \bibnamefont {Young}},\ }\href
  {\doibase 10.3390/educsci8010017} {\bibfield  {journal} {\bibinfo  {journal}
  {Education Sciences}\ }\textbf {\bibinfo {volume} {8}} (\bibinfo {year}
  {2018}),\ 10.3390/educsci8010017}\BibitemShut {NoStop}%
\bibitem [{\citenamefont {Arce-Larreta}\ \emph {et~al.}(2022)\citenamefont
  {Arce-Larreta}, \citenamefont {Assamagan}, \citenamefont {Barzi},
  \citenamefont {Bilow}, \citenamefont {Cecire}, \citenamefont {de~Jong},
  \citenamefont {Donati}, \citenamefont {Goldfarb}, \citenamefont {Klammer},
  \citenamefont {Muronga},\ and\ \citenamefont {Niland}}]{QuarknetHS}%
  \BibitemOpen
  \bibfield  {author} {\bibinfo {author} {\bibfnamefont {E.}~\bibnamefont
  {Arce-Larreta}}, \bibinfo {author} {\bibfnamefont {K.}~\bibnamefont
  {Assamagan}}, \bibinfo {author} {\bibfnamefont {E.}~\bibnamefont {Barzi}},
  \bibinfo {author} {\bibfnamefont {U.}~\bibnamefont {Bilow}}, \bibinfo
  {author} {\bibfnamefont {K.}~\bibnamefont {Cecire}}, \bibinfo {author}
  {\bibfnamefont {S.}~\bibnamefont {de~Jong}}, \bibinfo {author} {\bibfnamefont
  {S.}~\bibnamefont {Donati}}, \bibinfo {author} {\bibfnamefont
  {S.}~\bibnamefont {Goldfarb}}, \bibinfo {author} {\bibfnamefont
  {J.}~\bibnamefont {Klammer}}, \bibinfo {author} {\bibfnamefont
  {A.}~\bibnamefont {Muronga}}, \ and\ \bibinfo {author} {\bibfnamefont
  {M.}~\bibnamefont {Niland}},\ }\href@noop {} {\enquote {\bibinfo {title}
  {{The Necessity of International Particle Physics Opportunities for American
  Education}},}\ } (\bibinfo {year} {2022}),\ \Eprint
  {http://arxiv.org/abs/2203.09336} {arXiv:2203.09336 [physics.ed-ph]}
  \BibitemShut {NoStop}%
\bibitem [{\citenamefont {Dallal}\ \emph {et~al.}(2022)\citenamefont {Dallal},
  \citenamefont {Miller}, \citenamefont {Matten}, \citenamefont {Schur},
  \citenamefont {Sears}, \citenamefont {Carr}, \citenamefont {Rosenberg},
  \citenamefont {Unterman}, \citenamefont {Valsamis},\ and\ \citenamefont
  {Adams}}]{solar_eclipse}%
  \BibitemOpen
  \bibfield  {author} {\bibinfo {author} {\bibfnamefont {T.~A.}\ \bibnamefont
  {Dallal}}, \bibinfo {author} {\bibfnamefont {J.~M.}\ \bibnamefont {Miller}},
  \bibinfo {author} {\bibfnamefont {M.}~\bibnamefont {Matten}}, \bibinfo
  {author} {\bibfnamefont {E.}~\bibnamefont {Schur}}, \bibinfo {author}
  {\bibfnamefont {A.~J.}\ \bibnamefont {Sears}}, \bibinfo {author}
  {\bibfnamefont {C.}~\bibnamefont {Carr}}, \bibinfo {author} {\bibfnamefont
  {J.}~\bibnamefont {Rosenberg}}, \bibinfo {author} {\bibfnamefont {N.~A.}\
  \bibnamefont {Unterman}}, \bibinfo {author} {\bibfnamefont {A.}~\bibnamefont
  {Valsamis}}, \ and\ \bibinfo {author} {\bibfnamefont {M.}~\bibnamefont
  {Adams}},\ }\href {\doibase 10.1119/10.0009417} {\bibfield  {journal}
  {\bibinfo  {journal} {The Physics Teacher}\ }\textbf {\bibinfo {volume}
  {60}},\ \bibinfo {pages} {100} (\bibinfo {year} {2022})},\ \Eprint
  {http://arxiv.org/abs/https://doi.org/10.1119/10.0009417}
  {https://doi.org/10.1119/10.0009417} \BibitemShut {NoStop}%
\bibitem [{\citenamefont {Shaffer}(2010)}]{Shaffer_Quark}%
  \BibitemOpen
  \bibfield  {author} {\bibinfo {author} {\bibfnamefont {M.~D.}\ \bibnamefont
  {Shaffer}},\ }\emph {\bibinfo {title} {{The Experimentally Determined Average
  Flux Rate of Cosmic Ray Muons Near Topeka, Kansas}}},\ \href@noop {}
  {Master's thesis},\ \bibinfo  {school} {Emporia State University} (\bibinfo
  {year} {2010})\BibitemShut {NoStop}%
\bibitem [{\citenamefont {Gaisser}\ \emph {et~al.}(2016)\citenamefont
  {Gaisser}, \citenamefont {Engel},\ and\ \citenamefont
  {Resconi}}]{Gaisser2016}%
  \BibitemOpen
  \bibfield  {author} {\bibinfo {author} {\bibfnamefont {T.~K.}\ \bibnamefont
  {Gaisser}}, \bibinfo {author} {\bibfnamefont {R.}~\bibnamefont {Engel}}, \
  and\ \bibinfo {author} {\bibfnamefont {E.}~\bibnamefont {Resconi}},\
  }\href@noop {} {\emph {\bibinfo {title} {{Cosmic Rays and Particle
  Physics}}}}\ (\bibinfo  {publisher} {Cambridge University Press},\ \bibinfo
  {year} {2016})\BibitemShut {NoStop}%
\bibitem [{\citenamefont {Rao}\ and\ \citenamefont
  {Sreekantan}(1998)}]{rao_extensive_1998}%
  \BibitemOpen
  \bibfield  {author} {\bibinfo {author} {\bibfnamefont {M.~V.}\ \bibnamefont
  {Rao}}\ and\ \bibinfo {author} {\bibfnamefont {B.~V.}\ \bibnamefont
  {Sreekantan}},\ }\href@noop {} {\emph {\bibinfo {title} {{Extensive Air
  Showers}}}}\ (\bibinfo  {publisher} {World scientific},\ \bibinfo {year}
  {1998})\BibitemShut {NoStop}%
\bibitem [{\citenamefont {Tanabashi}\ \emph {et~al.}(2018)\citenamefont
  {Tanabashi}, \citenamefont {Hagiwara}, \citenamefont {Hikasa}, \citenamefont
  {Nakamura}, \citenamefont {Sumino}, \citenamefont {Takahashi}, \citenamefont
  {Tanaka}, \citenamefont {Agashe}, \citenamefont {Aielli}, \citenamefont
  {Amsler}, \citenamefont {Antonelli}, \citenamefont {Asner}, \citenamefont
  {Baer}, \citenamefont {Banerjee}, \citenamefont {Barnett}, \citenamefont
  {Basaglia}, \citenamefont {Bauer}, \citenamefont {Beatty}, \citenamefont
  {Belousov}, \citenamefont {Beringer}, \citenamefont {Bethke}, \citenamefont
  {Bettini}, \citenamefont {Bichsel}, \citenamefont {Biebel}, \citenamefont
  {Black}, \citenamefont {Blucher}, \citenamefont {Buchmuller}, \citenamefont
  {Burkert}, \citenamefont {Bychkov}, \citenamefont {Cahn}, \citenamefont
  {Carena}, \citenamefont {Ceccucci}, \citenamefont {Cerri}, \citenamefont
  {Chakraborty}, \citenamefont {Chen}, \citenamefont {Chivukula}, \citenamefont
  {Cowan}, \citenamefont {Dahl}, \citenamefont {D'Ambrosio}, \citenamefont
  {Damour}, \citenamefont {de~Florian}, \citenamefont {de~Gouv\^ea},
  \citenamefont {DeGrand}, \citenamefont {de~Jong}, \citenamefont {Dissertori},
  \citenamefont {Dobrescu}, \citenamefont {D'Onofrio}, \citenamefont {Doser},
  \citenamefont {Drees}, \citenamefont {Dreiner}, \citenamefont {Dwyer},
  \citenamefont {Eerola}, \citenamefont {Eidelman}, \citenamefont {Ellis},
  \citenamefont {Erler}, \citenamefont {Ezhela}, \citenamefont {Fetscher},
  \citenamefont {Fields}, \citenamefont {Firestone}, \citenamefont {Foster},
  \citenamefont {Freitas}, \citenamefont {Gallagher}, \citenamefont {Garren},
  \citenamefont {Gerber}, \citenamefont {Gerbier}, \citenamefont {Gershon},
  \citenamefont {Gershtein}, \citenamefont {Gherghetta}, \citenamefont
  {Godizov}, \citenamefont {Goodman}, \citenamefont {Grab}, \citenamefont
  {Gritsan}, \citenamefont {Grojean}, \citenamefont {Groom}, \citenamefont
  {Gr\"unewald}, \citenamefont {Gurtu}, \citenamefont {Gutsche}, \citenamefont
  {Haber}, \citenamefont {Hanhart}, \citenamefont {Hashimoto}, \citenamefont
  {Hayato}, \citenamefont {Hayes}, \citenamefont {Hebecker}, \citenamefont
  {Heinemeyer}, \citenamefont {Heltsley}, \citenamefont {Hern\'andez-Rey},
  \citenamefont {Hisano}, \citenamefont {H\"ocker}, \citenamefont {Holder},
  \citenamefont {Holtkamp}, \citenamefont {Hyodo}, \citenamefont {Irwin},
  \citenamefont {Johnson}, \citenamefont {Kado}, \citenamefont {Karliner},
  \citenamefont {Katz}, \citenamefont {Klein}, \citenamefont {Klempt},
  \citenamefont {Kowalewski}, \citenamefont {Krauss}, \citenamefont {Kreps},
  \citenamefont {Krusche}, \citenamefont {Kuyanov}, \citenamefont {Kwon},
  \citenamefont {Lahav}, \citenamefont {Laiho}, \citenamefont {Lesgourgues},
  \citenamefont {Liddle}, \citenamefont {Ligeti}, \citenamefont {Lin},
  \citenamefont {Lippmann}, \citenamefont {Liss}, \citenamefont {Littenberg},
  \citenamefont {Lugovsky}, \citenamefont {Lugovsky}, \citenamefont {Lusiani},
  \citenamefont {Makida}, \citenamefont {Maltoni}, \citenamefont {Mannel},
  \citenamefont {Manohar}, \citenamefont {Marciano}, \citenamefont {Martin},
  \citenamefont {Masoni}, \citenamefont {Matthews}, \citenamefont
  {Mei\ss{}ner}, \citenamefont {Milstead}, \citenamefont {Mitchell},
  \citenamefont {M\"onig}, \citenamefont {Molaro}, \citenamefont {Moortgat},
  \citenamefont {Moskovic}, \citenamefont {Murayama}, \citenamefont {Narain},
  \citenamefont {Nason}, \citenamefont {Navas}, \citenamefont {Neubert},
  \citenamefont {Nevski}, \citenamefont {Nir}, \citenamefont {Olive},
  \citenamefont {Pagan~Griso}, \citenamefont {Parsons}, \citenamefont
  {Patrignani}, \citenamefont {Peacock}, \citenamefont {Pennington},
  \citenamefont {Petcov}, \citenamefont {Petrov}, \citenamefont {Pianori},
  \citenamefont {Piepke}, \citenamefont {Pomarol}, \citenamefont {Quadt},
  \citenamefont {Rademacker}, \citenamefont {Raffelt}, \citenamefont
  {Ratcliff}, \citenamefont {Richardson}, \citenamefont {Ringwald},
  \citenamefont {Roesler}, \citenamefont {Rolli}, \citenamefont {Romaniouk},
  \citenamefont {Rosenberg}, \citenamefont {Rosner}, \citenamefont {Rybka},
  \citenamefont {Ryutin}, \citenamefont {Sachrajda}, \citenamefont {Sakai},
  \citenamefont {Salam}, \citenamefont {Sarkar}, \citenamefont {Sauli},
  \citenamefont {Schneider}, \citenamefont {Scholberg}, \citenamefont
  {Schwartz}, \citenamefont {Scott}, \citenamefont {Sharma}, \citenamefont
  {Sharpe}, \citenamefont {Shutt}, \citenamefont {Silari}, \citenamefont
  {Sj\"ostrand}, \citenamefont {Skands}, \citenamefont {Skwarnicki},
  \citenamefont {Smith}, \citenamefont {Smoot}, \citenamefont {Spanier},
  \citenamefont {Spieler}, \citenamefont {Spiering}, \citenamefont {Stahl},
  \citenamefont {Stone}, \citenamefont {Sumiyoshi}, \citenamefont {Syphers},
  \citenamefont {Terashi}, \citenamefont {Terning}, \citenamefont {Thoma},
  \citenamefont {Thorne}, \citenamefont {Tiator}, \citenamefont {Titov},
  \citenamefont {Tkachenko}, \citenamefont {T\"ornqvist}, \citenamefont
  {Tovey}, \citenamefont {Valencia}, \citenamefont {Van~de Water},
  \citenamefont {Varelas}, \citenamefont {Venanzoni}, \citenamefont {Verde},
  \citenamefont {Vincter}, \citenamefont {Vogel}, \citenamefont {Vogt},
  \citenamefont {Wakely}, \citenamefont {Walkowiak}, \citenamefont {Walter},
  \citenamefont {Wands}, \citenamefont {Ward}, \citenamefont {Wascko},
  \citenamefont {Weiglein}, \citenamefont {Weinberg}, \citenamefont {Weinberg},
  \citenamefont {White}, \citenamefont {Wiencke}, \citenamefont {Willocq},
  \citenamefont {Wohl}, \citenamefont {Womersley}, \citenamefont {Woody},
  \citenamefont {Workman}, \citenamefont {Yao}, \citenamefont {Zeller},
  \citenamefont {Zenin}, \citenamefont {Zhu}, \citenamefont {Zhu},
  \citenamefont {Zimmermann}, \citenamefont {Zyla}, \citenamefont {Anderson},
  \citenamefont {Fuller}, \citenamefont {Lugovsky},\ and\ \citenamefont
  {Schaffner}}]{constants}%
  \BibitemOpen
  \bibfield  {author} {\bibinfo {author} {\bibfnamefont {M.}~\bibnamefont
  {Tanabashi}}, \bibinfo {author} {\bibfnamefont {K.}~\bibnamefont {Hagiwara}},
  \bibinfo {author} {\bibfnamefont {K.}~\bibnamefont {Hikasa}}, \bibinfo
  {author} {\bibfnamefont {K.}~\bibnamefont {Nakamura}}, \bibinfo {author}
  {\bibfnamefont {Y.}~\bibnamefont {Sumino}}, \bibinfo {author} {\bibfnamefont
  {F.}~\bibnamefont {Takahashi}}, \bibinfo {author} {\bibfnamefont
  {J.}~\bibnamefont {Tanaka}}, \bibinfo {author} {\bibfnamefont
  {K.}~\bibnamefont {Agashe}}, \bibinfo {author} {\bibfnamefont
  {G.}~\bibnamefont {Aielli}}, \bibinfo {author} {\bibfnamefont
  {C.}~\bibnamefont {Amsler}}, \bibinfo {author} {\bibfnamefont
  {M.}~\bibnamefont {Antonelli}}, \bibinfo {author} {\bibfnamefont {D.~M.}\
  \bibnamefont {Asner}}, \bibinfo {author} {\bibfnamefont {H.}~\bibnamefont
  {Baer}}, \bibinfo {author} {\bibfnamefont {S.}~\bibnamefont {Banerjee}},
  \bibinfo {author} {\bibfnamefont {R.~M.}\ \bibnamefont {Barnett}}, \bibinfo
  {author} {\bibfnamefont {T.}~\bibnamefont {Basaglia}}, \bibinfo {author}
  {\bibfnamefont {C.~W.}\ \bibnamefont {Bauer}}, \bibinfo {author}
  {\bibfnamefont {J.~J.}\ \bibnamefont {Beatty}}, \bibinfo {author}
  {\bibfnamefont {V.~I.}\ \bibnamefont {Belousov}}, \bibinfo {author}
  {\bibfnamefont {J.}~\bibnamefont {Beringer}}, \bibinfo {author}
  {\bibfnamefont {S.}~\bibnamefont {Bethke}}, \bibinfo {author} {\bibfnamefont
  {A.}~\bibnamefont {Bettini}}, \bibinfo {author} {\bibfnamefont
  {H.}~\bibnamefont {Bichsel}}, \bibinfo {author} {\bibfnamefont
  {O.}~\bibnamefont {Biebel}}, \bibinfo {author} {\bibfnamefont {K.~M.}\
  \bibnamefont {Black}}, \bibinfo {author} {\bibfnamefont {E.}~\bibnamefont
  {Blucher}}, \bibinfo {author} {\bibfnamefont {O.}~\bibnamefont {Buchmuller}},
  \bibinfo {author} {\bibfnamefont {V.}~\bibnamefont {Burkert}}, \bibinfo
  {author} {\bibfnamefont {M.~A.}\ \bibnamefont {Bychkov}}, \bibinfo {author}
  {\bibfnamefont {R.~N.}\ \bibnamefont {Cahn}}, \bibinfo {author}
  {\bibfnamefont {M.}~\bibnamefont {Carena}}, \bibinfo {author} {\bibfnamefont
  {A.}~\bibnamefont {Ceccucci}}, \bibinfo {author} {\bibfnamefont
  {A.}~\bibnamefont {Cerri}}, \bibinfo {author} {\bibfnamefont
  {D.}~\bibnamefont {Chakraborty}}, \bibinfo {author} {\bibfnamefont {M.-C.}\
  \bibnamefont {Chen}}, \bibinfo {author} {\bibfnamefont {R.~S.}\ \bibnamefont
  {Chivukula}}, \bibinfo {author} {\bibfnamefont {G.}~\bibnamefont {Cowan}},
  \bibinfo {author} {\bibfnamefont {O.}~\bibnamefont {Dahl}}, \bibinfo {author}
  {\bibfnamefont {G.}~\bibnamefont {D'Ambrosio}}, \bibinfo {author}
  {\bibfnamefont {T.}~\bibnamefont {Damour}}, \bibinfo {author} {\bibfnamefont
  {D.}~\bibnamefont {de~Florian}}, \bibinfo {author} {\bibfnamefont
  {A.}~\bibnamefont {de~Gouv\^ea}}, \bibinfo {author} {\bibfnamefont
  {T.}~\bibnamefont {DeGrand}}, \bibinfo {author} {\bibfnamefont
  {P.}~\bibnamefont {de~Jong}}, \bibinfo {author} {\bibfnamefont
  {G.}~\bibnamefont {Dissertori}}, \bibinfo {author} {\bibfnamefont {B.~A.}\
  \bibnamefont {Dobrescu}}, \bibinfo {author} {\bibfnamefont {M.}~\bibnamefont
  {D'Onofrio}}, \bibinfo {author} {\bibfnamefont {M.}~\bibnamefont {Doser}},
  \bibinfo {author} {\bibfnamefont {M.}~\bibnamefont {Drees}}, \bibinfo
  {author} {\bibfnamefont {H.~K.}\ \bibnamefont {Dreiner}}, \bibinfo {author}
  {\bibfnamefont {D.~A.}\ \bibnamefont {Dwyer}}, \bibinfo {author}
  {\bibfnamefont {P.}~\bibnamefont {Eerola}}, \bibinfo {author} {\bibfnamefont
  {S.}~\bibnamefont {Eidelman}}, \bibinfo {author} {\bibfnamefont
  {J.}~\bibnamefont {Ellis}}, \bibinfo {author} {\bibfnamefont
  {J.}~\bibnamefont {Erler}}, \bibinfo {author} {\bibfnamefont {V.~V.}\
  \bibnamefont {Ezhela}}, \bibinfo {author} {\bibfnamefont {W.}~\bibnamefont
  {Fetscher}}, \bibinfo {author} {\bibfnamefont {B.~D.}\ \bibnamefont
  {Fields}}, \bibinfo {author} {\bibfnamefont {R.}~\bibnamefont {Firestone}},
  \bibinfo {author} {\bibfnamefont {B.}~\bibnamefont {Foster}}, \bibinfo
  {author} {\bibfnamefont {A.}~\bibnamefont {Freitas}}, \bibinfo {author}
  {\bibfnamefont {H.}~\bibnamefont {Gallagher}}, \bibinfo {author}
  {\bibfnamefont {L.}~\bibnamefont {Garren}}, \bibinfo {author} {\bibfnamefont
  {H.-J.}\ \bibnamefont {Gerber}}, \bibinfo {author} {\bibfnamefont
  {G.}~\bibnamefont {Gerbier}}, \bibinfo {author} {\bibfnamefont
  {T.}~\bibnamefont {Gershon}}, \bibinfo {author} {\bibfnamefont
  {Y.}~\bibnamefont {Gershtein}}, \bibinfo {author} {\bibfnamefont
  {T.}~\bibnamefont {Gherghetta}}, \bibinfo {author} {\bibfnamefont {A.~A.}\
  \bibnamefont {Godizov}}, \bibinfo {author} {\bibfnamefont {M.}~\bibnamefont
  {Goodman}}, \bibinfo {author} {\bibfnamefont {C.}~\bibnamefont {Grab}},
  \bibinfo {author} {\bibfnamefont {A.~V.}\ \bibnamefont {Gritsan}}, \bibinfo
  {author} {\bibfnamefont {C.}~\bibnamefont {Grojean}}, \bibinfo {author}
  {\bibfnamefont {D.~E.}\ \bibnamefont {Groom}}, \bibinfo {author}
  {\bibfnamefont {M.}~\bibnamefont {Gr\"unewald}}, \bibinfo {author}
  {\bibfnamefont {A.}~\bibnamefont {Gurtu}}, \bibinfo {author} {\bibfnamefont
  {T.}~\bibnamefont {Gutsche}}, \bibinfo {author} {\bibfnamefont {H.~E.}\
  \bibnamefont {Haber}}, \bibinfo {author} {\bibfnamefont {C.}~\bibnamefont
  {Hanhart}}, \bibinfo {author} {\bibfnamefont {S.}~\bibnamefont {Hashimoto}},
  \bibinfo {author} {\bibfnamefont {Y.}~\bibnamefont {Hayato}}, \bibinfo
  {author} {\bibfnamefont {K.~G.}\ \bibnamefont {Hayes}}, \bibinfo {author}
  {\bibfnamefont {A.}~\bibnamefont {Hebecker}}, \bibinfo {author}
  {\bibfnamefont {S.}~\bibnamefont {Heinemeyer}}, \bibinfo {author}
  {\bibfnamefont {B.}~\bibnamefont {Heltsley}}, \bibinfo {author}
  {\bibfnamefont {J.~J.}\ \bibnamefont {Hern\'andez-Rey}}, \bibinfo {author}
  {\bibfnamefont {J.}~\bibnamefont {Hisano}}, \bibinfo {author} {\bibfnamefont
  {A.}~\bibnamefont {H\"ocker}}, \bibinfo {author} {\bibfnamefont
  {J.}~\bibnamefont {Holder}}, \bibinfo {author} {\bibfnamefont
  {A.}~\bibnamefont {Holtkamp}}, \bibinfo {author} {\bibfnamefont
  {T.}~\bibnamefont {Hyodo}}, \bibinfo {author} {\bibfnamefont {K.~D.}\
  \bibnamefont {Irwin}}, \bibinfo {author} {\bibfnamefont {K.~F.}\ \bibnamefont
  {Johnson}}, \bibinfo {author} {\bibfnamefont {M.}~\bibnamefont {Kado}},
  \bibinfo {author} {\bibfnamefont {M.}~\bibnamefont {Karliner}}, \bibinfo
  {author} {\bibfnamefont {U.~F.}\ \bibnamefont {Katz}}, \bibinfo {author}
  {\bibfnamefont {S.~R.}\ \bibnamefont {Klein}}, \bibinfo {author}
  {\bibfnamefont {E.}~\bibnamefont {Klempt}}, \bibinfo {author} {\bibfnamefont
  {R.~V.}\ \bibnamefont {Kowalewski}}, \bibinfo {author} {\bibfnamefont
  {F.}~\bibnamefont {Krauss}}, \bibinfo {author} {\bibfnamefont
  {M.}~\bibnamefont {Kreps}}, \bibinfo {author} {\bibfnamefont
  {B.}~\bibnamefont {Krusche}}, \bibinfo {author} {\bibfnamefont {Y.~V.}\
  \bibnamefont {Kuyanov}}, \bibinfo {author} {\bibfnamefont {Y.}~\bibnamefont
  {Kwon}}, \bibinfo {author} {\bibfnamefont {O.}~\bibnamefont {Lahav}},
  \bibinfo {author} {\bibfnamefont {J.}~\bibnamefont {Laiho}}, \bibinfo
  {author} {\bibfnamefont {J.}~\bibnamefont {Lesgourgues}}, \bibinfo {author}
  {\bibfnamefont {A.}~\bibnamefont {Liddle}}, \bibinfo {author} {\bibfnamefont
  {Z.}~\bibnamefont {Ligeti}}, \bibinfo {author} {\bibfnamefont {C.-J.}\
  \bibnamefont {Lin}}, \bibinfo {author} {\bibfnamefont {C.}~\bibnamefont
  {Lippmann}}, \bibinfo {author} {\bibfnamefont {T.~M.}\ \bibnamefont {Liss}},
  \bibinfo {author} {\bibfnamefont {L.}~\bibnamefont {Littenberg}}, \bibinfo
  {author} {\bibfnamefont {K.~S.}\ \bibnamefont {Lugovsky}}, \bibinfo {author}
  {\bibfnamefont {S.~B.}\ \bibnamefont {Lugovsky}}, \bibinfo {author}
  {\bibfnamefont {A.}~\bibnamefont {Lusiani}}, \bibinfo {author} {\bibfnamefont
  {Y.}~\bibnamefont {Makida}}, \bibinfo {author} {\bibfnamefont
  {F.}~\bibnamefont {Maltoni}}, \bibinfo {author} {\bibfnamefont
  {T.}~\bibnamefont {Mannel}}, \bibinfo {author} {\bibfnamefont {A.~V.}\
  \bibnamefont {Manohar}}, \bibinfo {author} {\bibfnamefont {W.~J.}\
  \bibnamefont {Marciano}}, \bibinfo {author} {\bibfnamefont {A.~D.}\
  \bibnamefont {Martin}}, \bibinfo {author} {\bibfnamefont {A.}~\bibnamefont
  {Masoni}}, \bibinfo {author} {\bibfnamefont {J.}~\bibnamefont {Matthews}},
  \bibinfo {author} {\bibfnamefont {U.-G.}\ \bibnamefont {Mei\ss{}ner}},
  \bibinfo {author} {\bibfnamefont {D.}~\bibnamefont {Milstead}}, \bibinfo
  {author} {\bibfnamefont {R.~E.}\ \bibnamefont {Mitchell}}, \bibinfo {author}
  {\bibfnamefont {K.}~\bibnamefont {M\"onig}}, \bibinfo {author} {\bibfnamefont
  {P.}~\bibnamefont {Molaro}}, \bibinfo {author} {\bibfnamefont
  {F.}~\bibnamefont {Moortgat}}, \bibinfo {author} {\bibfnamefont
  {M.}~\bibnamefont {Moskovic}}, \bibinfo {author} {\bibfnamefont
  {H.}~\bibnamefont {Murayama}}, \bibinfo {author} {\bibfnamefont
  {M.}~\bibnamefont {Narain}}, \bibinfo {author} {\bibfnamefont
  {P.}~\bibnamefont {Nason}}, \bibinfo {author} {\bibfnamefont
  {S.}~\bibnamefont {Navas}}, \bibinfo {author} {\bibfnamefont
  {M.}~\bibnamefont {Neubert}}, \bibinfo {author} {\bibfnamefont
  {P.}~\bibnamefont {Nevski}}, \bibinfo {author} {\bibfnamefont
  {Y.}~\bibnamefont {Nir}}, \bibinfo {author} {\bibfnamefont {K.~A.}\
  \bibnamefont {Olive}}, \bibinfo {author} {\bibfnamefont {S.}~\bibnamefont
  {Pagan~Griso}}, \bibinfo {author} {\bibfnamefont {J.}~\bibnamefont
  {Parsons}}, \bibinfo {author} {\bibfnamefont {C.}~\bibnamefont {Patrignani}},
  \bibinfo {author} {\bibfnamefont {J.~A.}\ \bibnamefont {Peacock}}, \bibinfo
  {author} {\bibfnamefont {M.}~\bibnamefont {Pennington}}, \bibinfo {author}
  {\bibfnamefont {S.~T.}\ \bibnamefont {Petcov}}, \bibinfo {author}
  {\bibfnamefont {V.~A.}\ \bibnamefont {Petrov}}, \bibinfo {author}
  {\bibfnamefont {E.}~\bibnamefont {Pianori}}, \bibinfo {author} {\bibfnamefont
  {A.}~\bibnamefont {Piepke}}, \bibinfo {author} {\bibfnamefont
  {A.}~\bibnamefont {Pomarol}}, \bibinfo {author} {\bibfnamefont
  {A.}~\bibnamefont {Quadt}}, \bibinfo {author} {\bibfnamefont
  {J.}~\bibnamefont {Rademacker}}, \bibinfo {author} {\bibfnamefont
  {G.}~\bibnamefont {Raffelt}}, \bibinfo {author} {\bibfnamefont {B.~N.}\
  \bibnamefont {Ratcliff}}, \bibinfo {author} {\bibfnamefont {P.}~\bibnamefont
  {Richardson}}, \bibinfo {author} {\bibfnamefont {A.}~\bibnamefont
  {Ringwald}}, \bibinfo {author} {\bibfnamefont {S.}~\bibnamefont {Roesler}},
  \bibinfo {author} {\bibfnamefont {S.}~\bibnamefont {Rolli}}, \bibinfo
  {author} {\bibfnamefont {A.}~\bibnamefont {Romaniouk}}, \bibinfo {author}
  {\bibfnamefont {L.~J.}\ \bibnamefont {Rosenberg}}, \bibinfo {author}
  {\bibfnamefont {J.~L.}\ \bibnamefont {Rosner}}, \bibinfo {author}
  {\bibfnamefont {G.}~\bibnamefont {Rybka}}, \bibinfo {author} {\bibfnamefont
  {R.~A.}\ \bibnamefont {Ryutin}}, \bibinfo {author} {\bibfnamefont {C.~T.}\
  \bibnamefont {Sachrajda}}, \bibinfo {author} {\bibfnamefont {Y.}~\bibnamefont
  {Sakai}}, \bibinfo {author} {\bibfnamefont {G.~P.}\ \bibnamefont {Salam}},
  \bibinfo {author} {\bibfnamefont {S.}~\bibnamefont {Sarkar}}, \bibinfo
  {author} {\bibfnamefont {F.}~\bibnamefont {Sauli}}, \bibinfo {author}
  {\bibfnamefont {O.}~\bibnamefont {Schneider}}, \bibinfo {author}
  {\bibfnamefont {K.}~\bibnamefont {Scholberg}}, \bibinfo {author}
  {\bibfnamefont {A.~J.}\ \bibnamefont {Schwartz}}, \bibinfo {author}
  {\bibfnamefont {D.}~\bibnamefont {Scott}}, \bibinfo {author} {\bibfnamefont
  {V.}~\bibnamefont {Sharma}}, \bibinfo {author} {\bibfnamefont {S.~R.}\
  \bibnamefont {Sharpe}}, \bibinfo {author} {\bibfnamefont {T.}~\bibnamefont
  {Shutt}}, \bibinfo {author} {\bibfnamefont {M.}~\bibnamefont {Silari}},
  \bibinfo {author} {\bibfnamefont {T.}~\bibnamefont {Sj\"ostrand}}, \bibinfo
  {author} {\bibfnamefont {P.}~\bibnamefont {Skands}}, \bibinfo {author}
  {\bibfnamefont {T.}~\bibnamefont {Skwarnicki}}, \bibinfo {author}
  {\bibfnamefont {J.~G.}\ \bibnamefont {Smith}}, \bibinfo {author}
  {\bibfnamefont {G.~F.}\ \bibnamefont {Smoot}}, \bibinfo {author}
  {\bibfnamefont {S.}~\bibnamefont {Spanier}}, \bibinfo {author} {\bibfnamefont
  {H.}~\bibnamefont {Spieler}}, \bibinfo {author} {\bibfnamefont
  {C.}~\bibnamefont {Spiering}}, \bibinfo {author} {\bibfnamefont
  {A.}~\bibnamefont {Stahl}}, \bibinfo {author} {\bibfnamefont {S.~L.}\
  \bibnamefont {Stone}}, \bibinfo {author} {\bibfnamefont {T.}~\bibnamefont
  {Sumiyoshi}}, \bibinfo {author} {\bibfnamefont {M.~J.}\ \bibnamefont
  {Syphers}}, \bibinfo {author} {\bibfnamefont {K.}~\bibnamefont {Terashi}},
  \bibinfo {author} {\bibfnamefont {J.}~\bibnamefont {Terning}}, \bibinfo
  {author} {\bibfnamefont {U.}~\bibnamefont {Thoma}}, \bibinfo {author}
  {\bibfnamefont {R.~S.}\ \bibnamefont {Thorne}}, \bibinfo {author}
  {\bibfnamefont {L.}~\bibnamefont {Tiator}}, \bibinfo {author} {\bibfnamefont
  {M.}~\bibnamefont {Titov}}, \bibinfo {author} {\bibfnamefont {N.~P.}\
  \bibnamefont {Tkachenko}}, \bibinfo {author} {\bibfnamefont {N.~A.}\
  \bibnamefont {T\"ornqvist}}, \bibinfo {author} {\bibfnamefont {D.~R.}\
  \bibnamefont {Tovey}}, \bibinfo {author} {\bibfnamefont {G.}~\bibnamefont
  {Valencia}}, \bibinfo {author} {\bibfnamefont {R.}~\bibnamefont {Van~de
  Water}}, \bibinfo {author} {\bibfnamefont {N.}~\bibnamefont {Varelas}},
  \bibinfo {author} {\bibfnamefont {G.}~\bibnamefont {Venanzoni}}, \bibinfo
  {author} {\bibfnamefont {L.}~\bibnamefont {Verde}}, \bibinfo {author}
  {\bibfnamefont {M.~G.}\ \bibnamefont {Vincter}}, \bibinfo {author}
  {\bibfnamefont {P.}~\bibnamefont {Vogel}}, \bibinfo {author} {\bibfnamefont
  {A.}~\bibnamefont {Vogt}}, \bibinfo {author} {\bibfnamefont {S.~P.}\
  \bibnamefont {Wakely}}, \bibinfo {author} {\bibfnamefont {W.}~\bibnamefont
  {Walkowiak}}, \bibinfo {author} {\bibfnamefont {C.~W.}\ \bibnamefont
  {Walter}}, \bibinfo {author} {\bibfnamefont {D.}~\bibnamefont {Wands}},
  \bibinfo {author} {\bibfnamefont {D.~R.}\ \bibnamefont {Ward}}, \bibinfo
  {author} {\bibfnamefont {M.~O.}\ \bibnamefont {Wascko}}, \bibinfo {author}
  {\bibfnamefont {G.}~\bibnamefont {Weiglein}}, \bibinfo {author}
  {\bibfnamefont {D.~H.}\ \bibnamefont {Weinberg}}, \bibinfo {author}
  {\bibfnamefont {E.~J.}\ \bibnamefont {Weinberg}}, \bibinfo {author}
  {\bibfnamefont {M.}~\bibnamefont {White}}, \bibinfo {author} {\bibfnamefont
  {L.~R.}\ \bibnamefont {Wiencke}}, \bibinfo {author} {\bibfnamefont
  {S.}~\bibnamefont {Willocq}}, \bibinfo {author} {\bibfnamefont {C.~G.}\
  \bibnamefont {Wohl}}, \bibinfo {author} {\bibfnamefont {J.}~\bibnamefont
  {Womersley}}, \bibinfo {author} {\bibfnamefont {C.~L.}\ \bibnamefont
  {Woody}}, \bibinfo {author} {\bibfnamefont {R.~L.}\ \bibnamefont {Workman}},
  \bibinfo {author} {\bibfnamefont {W.-M.}\ \bibnamefont {Yao}}, \bibinfo
  {author} {\bibfnamefont {G.~P.}\ \bibnamefont {Zeller}}, \bibinfo {author}
  {\bibfnamefont {O.~V.}\ \bibnamefont {Zenin}}, \bibinfo {author}
  {\bibfnamefont {R.-Y.}\ \bibnamefont {Zhu}}, \bibinfo {author} {\bibfnamefont
  {S.-L.}\ \bibnamefont {Zhu}}, \bibinfo {author} {\bibfnamefont
  {F.}~\bibnamefont {Zimmermann}}, \bibinfo {author} {\bibfnamefont {P.~A.}\
  \bibnamefont {Zyla}}, \bibinfo {author} {\bibfnamefont {J.}~\bibnamefont
  {Anderson}}, \bibinfo {author} {\bibfnamefont {L.}~\bibnamefont {Fuller}},
  \bibinfo {author} {\bibfnamefont {V.~S.}\ \bibnamefont {Lugovsky}}, \ and\
  \bibinfo {author} {\bibfnamefont {P.}~\bibnamefont {Schaffner}} (\bibinfo
  {collaboration} {Particle Data Group}),\ }\href {\doibase
  10.1103/PhysRevD.98.030001} {\bibfield  {journal} {\bibinfo  {journal} {Phys.
  Rev. D}\ }\textbf {\bibinfo {volume} {98}},\ \bibinfo {pages} {030001}
  (\bibinfo {year} {2018})}\BibitemShut {NoStop}%
\bibitem [{\citenamefont {Shukla}\ and\ \citenamefont
  {Sankrith}(2018)}]{shukla_energy_2018}%
  \BibitemOpen
  \bibfield  {author} {\bibinfo {author} {\bibfnamefont {P.}~\bibnamefont
  {Shukla}}\ and\ \bibinfo {author} {\bibfnamefont {S.}~\bibnamefont
  {Sankrith}},\ }\href
  {https://www.worldscientific.com/doi/abs/10.1142/S0217751X18501750}
  {\bibfield  {journal} {\bibinfo  {journal} {International Journal of Modern
  Physics A}\ }\textbf {\bibinfo {volume} {33}},\ \bibinfo {pages} {1850175}
  (\bibinfo {year} {2018})}\BibitemShut {NoStop}%
\bibitem [{\citenamefont {Schwerdt}(2018)}]{schwerdt_zenith_2018}%
  \BibitemOpen
  \bibfield  {author} {\bibinfo {author} {\bibfnamefont {C.}~\bibnamefont
  {Schwerdt}},\ }\href
  {https://icd.desy.de/sites/sites_conferences/site_icd/content/e12688/e13082/e80711/Cosmic@Web_engl.pdf}
  {\bibfield  {journal} {\bibinfo  {journal} {Wissenschaftliche Koordinatorin
  Cosmic-Projekte, Zeuthen}\ }\textbf {\bibinfo {volume} {21}} (\bibinfo {year}
  {2018})}\BibitemShut {NoStop}%
\bibitem [{\citenamefont {Pethuraj}\ \emph {et~al.}(2017)\citenamefont
  {Pethuraj}, \citenamefont {Datar}, \citenamefont {Majumder}, \citenamefont
  {Mondal}, \citenamefont {Ravindran},\ and\ \citenamefont
  {Satyanarayana}}]{Pethuraj_2017}%
  \BibitemOpen
  \bibfield  {author} {\bibinfo {author} {\bibfnamefont {S.}~\bibnamefont
  {Pethuraj}}, \bibinfo {author} {\bibfnamefont {V.}~\bibnamefont {Datar}},
  \bibinfo {author} {\bibfnamefont {G.}~\bibnamefont {Majumder}}, \bibinfo
  {author} {\bibfnamefont {N.}~\bibnamefont {Mondal}}, \bibinfo {author}
  {\bibfnamefont {K.}~\bibnamefont {Ravindran}}, \ and\ \bibinfo {author}
  {\bibfnamefont {B.}~\bibnamefont {Satyanarayana}},\ }\href {\doibase
  10.1088/1475-7516/2017/09/021} {\bibfield  {journal} {\bibinfo  {journal}
  {Journal of Cosmology and Astroparticle Physics}\ }\textbf {\bibinfo {volume}
  {2017}},\ \bibinfo {pages} {021} (\bibinfo {year} {2017})}\BibitemShut
  {NoStop}%
\bibitem [{\citenamefont {Fermilab}(2012)}]{eLab}%
  \BibitemOpen
  \bibfield  {author} {\bibinfo {author} {\bibnamefont {Fermilab}},\ }\href
  {https://www.i2u2.org/elab/cosmic/home/project.jsp} {\enquote {\bibinfo
  {title} {{Cosmic Ray e-Lab}},}\ }\bibinfo {howpublished} {Online} (\bibinfo
  {year} {2012})\BibitemShut {NoStop}%
\bibitem [{QUA(2012)}]{QUACRD_instructions}%
  \BibitemOpen
  \href
  {https://quarknet.org/sites/default/files/cf_crmdassemblyinstructions-small.pdf}
  {\emph {\bibinfo {title} {{QuarkNet Cosmic Ray Muon Detector (CRMD) Assembly
  Instructions for Series 6000 DAQ}}}} (\bibinfo {year} {2012})\BibitemShut
  {NoStop}%
\bibitem [{\citenamefont {Lofgren}(2001)}]{lofgren_quarknet_2001}%
  \BibitemOpen
  \bibfield  {author} {\bibinfo {author} {\bibfnamefont {J.}~\bibnamefont
  {Lofgren}},\ }\href
  {https://research.fit.edu/media/site-specific/researchfitedu/hep/quarknet/documents/Quarknet_card1_referenc.pdf}
  {\enquote {\bibinfo {title} {{Quarknet Cosmic Ray Detection System}},}\
  }\bibinfo {howpublished} {Online} (\bibinfo {year} {2001})\BibitemShut
  {NoStop}%
\bibitem [{\citenamefont {Fermilab}(2001{\natexlab{a}})}]{performance}%
  \BibitemOpen
  \bibfield  {author} {\bibinfo {author} {\bibnamefont {Fermilab}},\ }\href
  {http://www.i2u2.org/elab/cosmic/analysis-performance/tutorial.jsp} {\enquote
  {\bibinfo {title} {{Performance Study Tutorial}},}\ }\bibinfo {howpublished}
  {Online} (\bibinfo {year} {2001}{\natexlab{a}})\BibitemShut {NoStop}%
\bibitem [{\citenamefont {Fermilab}(2001{\natexlab{b}})}]{signal}%
  \BibitemOpen
  \bibfield  {author} {\bibinfo {author} {\bibnamefont {Fermilab}},\ }\href
  {http://www.i2u2.org/elab/cosmic/references/display.jsp?name=signal_width&type=glossary}
  {\enquote {\bibinfo {title} {{Signal Width}},}\ }\bibinfo {howpublished}
  {Online} (\bibinfo {year} {2001}{\natexlab{b}})\BibitemShut {NoStop}%
\bibitem [{pas(2009)}]{paschke_calibration_2009}%
  \BibitemOpen
  \href {http://faculty.ucr.edu/~ellison/Quarknet/6000CRMD_How_to_Plateau.pdf}
  {\enquote {\bibinfo {title} {{Calibration Instructions for Quarknet Cosmic
  Ray Detector}},}\ } (\bibinfo {year} {2009}),\ \bibinfo {note}
  {presentation}\BibitemShut {NoStop}%
\bibitem [{\citenamefont {McGraw-Hill}(1997)}]{steradian}%
  \BibitemOpen
  \bibfield  {author} {\bibinfo {author} {\bibnamefont {McGraw-Hill}},\
  }\href@noop {} {\emph {\bibinfo {title} {McGraw-Hill Dictionary of Scientific
  and Technical Terms}}},\ \bibinfo {edition} {fifth edition}\ ed.,\ edited by\
  \bibinfo {editor} {\bibfnamefont {S.~P.}\ \bibnamefont {Parker}}\ (\bibinfo
  {publisher} {McGraw-Hill Education},\ \bibinfo {year} {1997})\BibitemShut
  {NoStop}%
\bibitem [{\citenamefont {Bektasoglu}\ and\ \citenamefont
  {Arslan}(2013)}]{bektasoglu_investigation_2013}%
  \BibitemOpen
  \bibfield  {author} {\bibinfo {author} {\bibfnamefont {M.}~\bibnamefont
  {Bektasoglu}}\ and\ \bibinfo {author} {\bibfnamefont {H.}~\bibnamefont
  {Arslan}},\ }\href {http://link.springer.com/10.1007/s12043-013-0519-2}
  {\bibfield  {journal} {\bibinfo  {journal} {Pramana}\ }\textbf {\bibinfo
  {volume} {80}},\ \bibinfo {pages} {837} (\bibinfo {year} {2013})}\BibitemShut
  {NoStop}%
\bibitem [{\citenamefont {Shteinbuk}(2011)}]{shteinbuk_measuring_nodate}%
  \BibitemOpen
  \bibfield  {author} {\bibinfo {author} {\bibfnamefont {I.}~\bibnamefont
  {Shteinbuk}},\ }\href
  {https://www.s.u-tokyo.ac.jp/en/utrip/archive/2011/pdf/17Inna.pdf} {\enquote
  {\bibinfo {title} {{Measuring the Angular Distribution of Muons}},}\ }
  (\bibinfo {year} {2011}),\ \bibinfo {note} {unpublished}\BibitemShut
  {NoStop}%
\bibitem [{\citenamefont {Bae}\ and\ \citenamefont {Chatzidakis}(2022)}]{Bae}%
  \BibitemOpen
  \bibfield  {author} {\bibinfo {author} {\bibfnamefont {J.}~\bibnamefont
  {Bae}}\ and\ \bibinfo {author} {\bibfnamefont {S.}~\bibnamefont
  {Chatzidakis}},\ }\href@noop {} {\enquote {\bibinfo {title} {{A New
  Semi-Empirical Model for Cosmic Ray Muon Flux Estimation}},}\ } (\bibinfo
  {year} {2022}),\ \Eprint {http://arxiv.org/abs/2110.14152} {arXiv:2110.14152
  [astro-ph.IM]} \BibitemShut {NoStop}%
\bibitem [{\citenamefont {Grieder}(2001)}]{grieder_cosmic_2001}%
  \BibitemOpen
  \bibfield  {author} {\bibinfo {author} {\bibfnamefont {P.~K.}\ \bibnamefont
  {Grieder}},\ }\href@noop {} {\emph {\bibinfo {title} {{Cosmic Rays at
  Earth}}}}\ (\bibinfo  {publisher} {Elsevier},\ \bibinfo {year}
  {2001})\BibitemShut {NoStop}%
\bibitem [{\citenamefont {Parra}\ and\ \citenamefont
  {Bernal}(2019)}]{useche_estimation_2019}%
  \BibitemOpen
  \bibfield  {author} {\bibinfo {author} {\bibfnamefont {J.~U.}\ \bibnamefont
  {Parra}}\ and\ \bibinfo {author} {\bibfnamefont {C.~{\'A}.}\ \bibnamefont
  {Bernal}},\ }\href {\doibase 10.1088/1748-0221/14/02/P02015} {\bibfield
  {journal} {\bibinfo  {journal} {Journal of Instrumentation}\ }\textbf
  {\bibinfo {volume} {14}},\ \bibinfo {pages} {P02015} (\bibinfo {year}
  {2019})}\BibitemShut {NoStop}%
\bibitem [{\citenamefont {Y{\'a}{\~n}ez}\ and\ \citenamefont
  {Aguilar-Arevalo}(2021)}]{Olmos_Y_ez_2021}%
  \BibitemOpen
  \bibfield  {author} {\bibinfo {author} {\bibfnamefont {B.~O.}\ \bibnamefont
  {Y{\'a}{\~n}ez}}\ and\ \bibinfo {author} {\bibfnamefont {A.~A.}\ \bibnamefont
  {Aguilar-Arevalo}},\ }\href
  {https://www.sciencedirect.com/science/article/pii/S0168900220312675}
  {\bibfield  {journal} {\bibinfo  {journal} {Nuclear Instruments and Methods
  in Physics Research Section A: Accelerators, Spectrometers, Detectors and
  Associated Equipment}\ }\textbf {\bibinfo {volume} {987}},\ \bibinfo {pages}
  {164870} (\bibinfo {year} {2021})}\BibitemShut {NoStop}%
\end{thebibliography}%

\end{document}